\documentclass [11pt]{article}
\usepackage[dvips]{color}
\usepackage{epsfig}
\usepackage[normalem]{ulem}

\def\beq{\begin{equation}}
\def\eeq{\end{equation}}
\def\beqa{\begin{eqnarray}}
\def\eeqa{\end{eqnarray}}
\def\d{{\rm d}}

\begin{document}
\baselineskip0.7cm plus 1pt minus 1pt
\tolerance=1500

\begin{center}
{\LARGE\bf Dissipation effects in mechanics and  thermodynamics}
\vskip0.4cm
{ J. G\"u\'emez$^{a,}$\footnote{guemezj@unican.es},
M. Fiolhais$^{b,}$\footnote{tmanuel@teor.fis.uc.pt}
}
\vskip0.1cm
{\it $^a$ Department of Applied Physics,  University of Cantabria,  Spain} \\
\vskip0.1cm
{\it $^b$ CFisUC, Department of Physics,  University of Coimbra, Portugal}
\vskip0.1cm
\end{center}

\begin{abstract}
With the discussion of three examples, we aim at clarifying the concept of energy transfer associated with  {\color{black} dissipation} in mechanics and in thermodynamics.
The {\color{black} dissipation effects due to} dissipative forces, such as the friction force between solids or the drag force in motions in fluids {\color{black}, lead to an internal energy increase of the system and/or to a heat transfer to the surrounding.}  This {\color{black} heat flow} is  {\color{black} consistent with} the second law, which states that
the entropy of the universe should increase when those forces are present because of the irreversibility always associated with their actions. As far as mechanics is concerned the effects of the dissipative forces are include in the Newton's equations as impulses and pseudo-works.
\end{abstract}

\section{Introduction}

Many textbook exercises, presented in the context of mechanics, involve the action of dissipative forces, therefore immediately entering the perimeter of thermodynamics \cite{walker11}.
In this paper we address three such classroom problems to highlight the need for using the laws of thermodynamics, besides the laws of mechanics, if one really wants to describe  the energetic aspects whenever we are in the presence of  dissipative  forces.

A very common approach to deal with textbook exercises of classical mechanics in secondary school and in
a typical general physics curricular unit is the reduction of the actual body to a so-called material point \cite{jewett08a}.
Although the representation of an extended body by a particle (its centre-of-mass) may have obvious advantages, that simplification usually
 does not always  allow  for a complete
analysis of the pertinent energetic issues, for instance, when  dissipative  forces are present.
This is true even when we are in the presence of just a translational motion.

A major point in this article is the discussion around the work \cite{mallinckrodt92}.
In mechanics, work is defined as a ``force---displacement" product \cite{hilborn00}.
The same is true in thermodynamics where the notion is {\color{black} extended} to the ``generalized force---generalized displacement" product involving conjugate variables \cite{zemansky97}.
But, in thermodynamics, there is a significant supplementary aspect, that cannot be circumvented, related to the absolute necessity to introduce the concept of work reservoir \cite{anacleto10}.
This is in close analogy
with the need for introducing the concept of heat reservoir now in relation with the heat transfer to or from the system.
In a limited  strict sense, work is the energy exchanged between a system and a work reservoir such that the mentioned energy can be totally stored in an organized manner either in the system or in reservoirs. This organized   energy  (which is stored)   can be always recovered in reversible processes with no change of the entropy of the universe. A work reservoir, by definition, only performs reversible processes (similarly to the heat reservoir) without increasing the entropy of the universe (this is not the case for a heat reservoir).
However, the mechanical description of processes involving dissipative forces uses equations in which there are terms with the dimension of an energy, which are products of a force by a displacement but they cannot be considered as  work in the sense  referred to above.

In particular, these products enter in the centre-of-mass equation (equivalent to the Newton's second law), and are  called pseudo-works \cite{penchina78}.
They account for the kinetic energy variation of the center-of-mass of the system but,  in general, they  are not work.

Actually, from the first law of thermodynamics point of view those products may represent energy exchanged between the system and the surrounding but they should not be regarded as work (i.e. they are not work).
In fact, some force-displacement products are related to physical processes in which the system does not interact energetically with a work reservoir. Commonly, one refers to the product of a dissipative force by a characteristic displacement of the body as dissipative work. At a  more formal level, one may say that this happens whenever a product ``generalized force---generalized displacement" involves  macroscopic variables which do not serve to characterize the state of the system, i.e. when they are not state variables of the system.
Moreover, {\color{black} the equivalence between}  dissipative work {\color{black} and}  heat{\color{black}, established long ago after the famous Joule's wheel paddle experiment,} is crucial to explain the irreversibility, with an increase of the entropy of the universe, which always takes place whenever dissipative forces play a role in a process \cite{silverberg07}. {\color{black} That experiment established the ``mechanical equivalent of heat" an expression coined by James Joule himself.}

We illustrate these ideas using concrete and even popular examples such as the motion of sliding blocks under the action of friction forces \cite{sherwood84},
 the motion of an accelerating car also acted upon by a drag force produced by the air resistance \cite{erlichson77},
and the rotation/translation motion of a disc descending a slippery slope, therefore with kinetic frictional forces playing a role.

Since we are dealing with very fundamental concepts, to make our points very clear we need to start with a necessarily brief review of some basic aspects of mechanics and thermodynamics. This is carried on in section \ref{sec:mechthermo}. In sections 3 to 5   we analyse three situations  which are usually considered as mechanical problems involving dissipative forces --- we use the examples to illustrate how to consistently deal with the effects of these forces both in mechanics and in thermodynamics. Finally, in section \ref{sec:conclus} we draw the conclusions.

\section{Mechanics and thermodynamics - a brief review of the pertinent equations}
\label{sec:mechthermo}

From the energetic point of view associated with their effects, we may distinguish between several types of forces. For instance, a conservative force can be expressed as the  {\color{black} negative} of the gradient of a potential field   (gravitational, electrostatic, etc.) and, through its action, the system always interacts with a work reservoir. For a non-conservative but  non-dissipative force, which cannot be expressed as a gradient, there is no work reservoir: usually these forces, such {\color{black} as the constraint forces }  as the normal force acting on an object placed or moving on top of a surface,  do not perform any work.  Other examples are the tension force in a pendulum{\color{black}, also a constraint force,} or the static friction force on a car's wheel or on the rim of a rolling non-slippering disc. On the other hand, there are dissipative forces, necessarily non-conservative ones {\color{black} (hence, no work reservoir involved)}, whose action leads to the dissipation of mechanical energy.  These forces can  only  be applied to extensive bodies \cite{besson01} and they are  {\color{black} central} in the discussion along this paper.
In spite of the requirement that a dissipative force, such as the kinetic friction force, can only be applied to extensive bodies, one often uses the point-like approximation for the body in a pure mechanical approach to the problem. Of course this approximation introduces limitations in the physical treatment of the situation and sometimes it also does introduce inconsistencies with respect to the realistic approach.

Let us summarize some important results for a system of $N$ particles, each of which with mass $m_i$. We admit that the particles may interact by means of two body forces described by a potential energy, $V_{ij}$, which depends on the
 distance $r_{ij}=|\vec r_j -\vec r_i|$. In a given reference frame, say the lab frame, the total energy of the system is the sum of the kinetic and potential energies, i.e.
\beq
E_{\rm total} = {1 \over 2} \sum _i m_i v_i^2 + {1 \over 2} \sum _{i, j\not= i} V_{ij}^{\rm int}
\label{etoruy}
\eeq
(the second 1/2 is introduced to avoid doublecounting). The dummy summation indexes run from 1 to $N$. The centre-of-mass velocity  is defined by
\beq
\vec V_{\rm cm} = {1 \over M}  \sum _i m_i \vec v_i \, ,
\label{cmhj}
\eeq
where $  M =\sum _i m_i$ is the total mass of the system.

The velocities $\vec v_i$ can always be written as
\beq
\vec v_i= \vec V_{\rm cm}  + \vec v{\, '}_i\, ,
\label{veloc}
\eeq
where by $\vec v{\, '}_i$ we denote the velocity of particle $i$ in the centre-of-mass frame. If we insert (\ref{veloc}) into (\ref{etoruy}), one arrives at
\beq
E_{\rm total} = {1 \over 2} M V_{\rm cm}^2 + {1 \over 2} \sum _i m_i {v'}_i^2 + {1 \over 2} \sum _{i, j\not= i} V_{ij}^{\rm int}\, ,
\label{totyue}
\eeq
where we have already used the fact that $\sum _i m_i \vec v{\, '}_i=\vec 0$, i.e. the centre-of-mass velocity vanishes in the centre-of-mass frame.
The three terms on the right hand side of equation (\ref{totyue}) are: the kinetic energy of the centre-of-mass,  ${K}_{\rm cm}$; the kinetic energy of the system with respect to the centre-of-mass, $K^{\rm int}$ (internal kinetic energy); and the potential energy
associated with the interactions {\em inside} the system, $V^{\rm int}$ (internal potential energy). The last two terms can be aggregated into a single one, the internal energy of the system, $U=K^{\rm int}+V^{\rm int}$. Therefore the total energy of the system of $N$ particles can be written, in a given inertial reference frame, as the sum of two terms:
 \beq
 E_{\rm total} = K_{\rm cm}+U\, .
 \eeq
To describe several physical situations, in recent papers \cite{guemez13, guemez14}
 we have used this {\color{black} splitting} of the total energy of a system  whose validity is then quite general under the assumed conditions.
Hence, the internal energy can always be regarded as the energy of the system that is not the centre-of-mass kinetic energy, {\color{black} irrespective of the reference frame}.

\subsection{The first law of thermodynamics}
\label{ssec:firstlawth}

 The first law of thermodynamics, established in the middle of the 19th century, is a statement on the energy conservation. For a system that interacts with the surrounding, its total energy varies by the exact amount of energy transferred through the
 system boundary:   $\Delta E_{\rm total}= E_{\rm transf}$ \cite{erlichson84}.
 On the one hand, this energy transfer can be work performed by the external forces, $W^{\rm ext}$. Indeed, any external force acting upon the system, in general gives rise to an energy transfer to/from the system, and the total external work is given by
\beq
W^{\rm ext} =  \sum_j \int { {\vec F}^{\rm ext}_j}\cdot \ \d {\vec r}_j\, ,  \label{trabalhoe}
\eeq
where ${\vec F}^{\rm ext}_j$ represents each external force acting on the system, and $\d {\vec r}_j$ its own infinitesimal displacement.  In (\ref{trabalhoe}), the dummy  $j$ is used now to label these external forces (not to label particles as previously).
On the other hand, any energy transfer that cannot be associated with work, should be considered  as heat, $Q$.
Therefore, the energy conservation  can be expressed in the following form \cite{guemez13}:
\beq
\Delta {K}_{\rm cm} + \Delta { U} = W^{\rm ext}  \ + \ Q\, ,
\label{totale}
\eeq
which expresses the first law of thermodynamics.
The left hand side accounts for the system total energy variation and the right hand side refers to the energy that crosses the system's boundary.
Once the energy had crossed the system boundary, it may either change the kinetic energy of the system centre-of-mass --- the first term on the left hand side of (\ref{totale}) ---, or the internal energy of the system{\color{black}, or both}. Of course, equation (\ref{totale}) also allows for a direct conversion of $U$ into ${K}_{\rm cm}$ and vice-versa, even when there is no energy crossing the system's boundary  in processes permitted by the second law.

\subsection{Newton's second law}
\label{ssec:newtonseclaw}

Much before the formulation of the first law of thermodynamics, Isaac Newton developed the theory for describing the translational motion of a body. For a single particle, {\color{black} labeled $i$, belonging to a system of $N$ particles,} Newton's second law can be formulated by the equation
\beq
\vec f_i \, \d t = m_i \d \vec v_i \, ,
\label{new0}
\eeq
where $\vec f_i$ is the resultant force acting upon particle $i$. We can scalar multiply both terms of the previous equation by $\vec v_i$ and integrate between an initial and a final state.
This leads to the well-known work-energy theorem
\beq
\int \vec f_i  \cdot \d \vec r_i = \Delta k_i  \, ,
\label{onepai}
\eeq
where  $\d \vec r_i = \vec v_i \, \d t$ is the infinitesimal displacement of particle $i$ in the time interval $\d t$ and $k_i$ is its kinetic energy. The work performed by the resultant force acting on the particle is equal to its kinetic energy variation. The resultant force acting on the particle is the sum of the internal and external resultant forces, $f^{\rm int}_i+f^{\rm ext}_i$. This splitting allows us to write the left hand side of (\ref{onepai}) as the sum of an internal and  external work, namely
\beq
w^{\rm int}_i+ w^{\rm ext}_i =  \Delta k_i\, ,
\eeq
where $w^{\rm int}_i=\int \vec f^{\rm int}_i  \cdot \d \vec r_i$ and similarly for the resultant external force acting on particle $i$. If we now go to the macroscopic world,
summing up over all particles, and defining $\Delta K= \sum _i\,\Delta k_i$  one arrives at
{\color{black}
\beq
\Delta K=  W^{\rm int}+W^{\rm ext}\, .  
\label{senhora}
\eeq
}
 This equation, {\color{black} algebraically derived from Newton's second law and therefore totally equivalent to it}, is a work-energy theorem for the system of particles, stating that the total variation of kinetic energy of the system is equal the internal and external  works.
\color{black}

{\color{black} We should note that equation (\ref{senhora}) is usually not operative for a many particle system (such as a thermodynamical system) because, in general, one does not explicitly know neither the forces acting on each particle nor the respective displacement}.   We {\color{black} also} note that the left hand side of equation (\ref{senhora}) can still be decomposed as
$\Delta K= \Delta K_{\rm cm}+\Delta K^{\rm int}$ (K\"onig theorem), which are the first and the second terms on the right hand side of equation (\ref{totyue}).

There is another way to proceed from equation   (\ref{new0}). We may sum over all particles and then use expression (\ref{cmhj}) to conclude that
\beq
\vec F^{\rm ext} \, \d t = M \d \vec V_{\rm cm}\, .
\label{hjaoi}
\eeq
The sum over all forces acting on  all particles, $\sum_i \vec f_i= \vec F^{\rm ext}$, is the resultant of all external forces (conservative and non-conservative) acting on the system, since the resultant of the internal forces must vanish (Newton's third law), ${\vec F}^{\rm int}=\vec 0$. After scalar multiplying both sides of
(\ref{hjaoi}) by $\vec V_{\rm cm}$, one arrives, after integration to
\beq
\Delta K_{\rm cm} = \int  {\vec F}^{\rm ext} \cdot \d {\vec R}_{\rm cm}\, ,
\label{eq-3}
\eeq
where $\d \vec R_{\rm cm}$ is the infinitesimal displacement of the centre-of-mass. The force---displacement product ${\vec F}^{\rm ext} \cdot \d {\vec R}_{\rm cm}$ is usually called  pseudo-work \cite{sherwood83}
(it artificially becomes work if the body is assumed as a point-like object).

{\color{black} Though} equation (\ref{eq-3}) is a well-known result, at this point we should stress that {\color{black} it was algebraically derived from Newton's second law (plus the third law) applied to a system of particles, hence it is equivalent to that fundamental law.}  {\color{black} Both equations (\ref{senhora}) and (\ref{eq-3}) are different forms of expressing Newton's second law which is usually given in vector form such as  (\ref{hjaoi}).}

One should also  note that the right hand side in (\ref{eq-3}) is not the external work given by equation (\ref{trabalhoe}). In equation
(\ref{trabalhoe}) each external force and its own displacement is considered separately, whereas in  (\ref{eq-3}) it is the resultant of the external forces and the displacement of the centre-of-mass that appears. To emphasize the distinction, the right hand side of (\ref{eq-3}) is usually referred to as pseudo-work, but the name does not matter. The important point is to realize that, in general, $W^{\rm ext}\not= \Delta K _{\rm cm}$ \cite{bauman92}
or, more explicitly, in general $\sum_{j} \int  {\vec F}^{\rm ext}_j \cdot \ \d {\vec r}_j \not=
\int  {\vec F}^{\rm ext} \cdot \d {\vec R}_{\rm cm}$, where ${\vec F}^{\rm ext}=\sum_j {\vec F}^{\rm ext}_j$.

For a system with moment of inertia $I$, rotating around a principal axis {\color{black} the angular impulse---angular momentum  equation and the kinetic energy equation, corresponding  to  (\ref{hjaoi}) and (\ref{eq-3}) in the case of rotations} are given, in differential form, by \cite{guemez14a}
\beqa
\Gamma^{\rm ext} \, \d t &=& I\, \d \omega \label{rot1}\\
\d K^{\rm rot} & = &   \Gamma^{\rm ext} \d \theta   \, , \label{rot2}
\eeqa
where $\Gamma^{\rm ext} $ is the resultant external torque ($\Gamma^{\rm int}=0$), $\omega$ the angular velocity and $\theta$ the rotation angle. Of course, the rotational energy variation $\Delta K^{\rm rot}$
is part of $\Delta K^{\rm int}$ that enters in $\Delta U$.

In sections 3---5 we are going to study three examples illustrating the complementarity of the first law of thermodynamics {\color{black} with respect to Newton's second law.}

\subsection{Work in mechanics and in thermodynamics}
\label{ssec:workmecther}

Equations  (\ref{totale}) and (\ref{eq-3}) [or (\ref{senhora})], express the first law of thermodynamics and the Newton's second law, respectively, and are general, i.e. they are applicable to any system, of whatever nature.

In mechanics, the energy interaction is only accountable through  force---displacement products. But, in thermodynamics, besides work, the heat is another mechanism to account for an energetic interaction between the system and the surrounding.

There are some subtle points associated with the work even in the pure mechanical framework, when one goes from the real extended system to the description of its motion only in terms of its centre-of-mass \cite{bauman92}.
For instance, when one walks, the horizontal force  {\color{black} on} the foot, pointing alternately in the forward and in the backward directions, does not perform any work --- it is a static friction force whose application point does not move \cite{guemez13c}. Now, if one considers the body reduced to a single point, the horizontal force is applied at this point that moves and, therefore, there is indeed work associated with the force. The appearance of this work is then an artifact of the  {\color{black} particle model approach}  since we know that, for the real system, that force does not perform any work whatsoever. A force that is called ``static friction force" now moves its application point, contradicting its own denomination, all because of the  {\color{black} model} in use! The instructor should be very clear when he explains the ``work" of that force --- we even suggest to suppress the denomination ``static" force if the body is reduced to its centre-of-mass. The right hand side of equation (\ref{eq-3}) {\em is not} work but it (artificially) ``becomes" work if the extended body is reduced, from the outset, to a point-like object.

There are several identical examples of forces that do not perform any work if the extended (usually deformable) system is considered, but they seem to  do when the description of the motion reduces to the centre-of-mass.  In previous papers we discussed the physics  of a person that jumps \cite{guemez14}:
during the time interval the person keeps contact with the ground, the normal force on the feet does not do any work. But it does if the problem is discussed after reducing the person to a point-like object. A car in motion \cite{guemez13b}, to be discussed in section \ref{sec:motioncar}, provides another good example of the point we want to make:  the car accelerates in the forward direction due to the forward static friction forces exerted by the ground on the tires. These (neither conservative nor dissipative) forces do not perform any work (the car moves thanks to the gas consumption in the engine, there is no energy coming from the ground). The force exerted by the infinite mass body (the ground) intermediates the process of converting internal energy into kinetic energy. However, if the car is reduced to a single moving point, the tractive forces do work indeed, though this work is again an artifact of the point-like  {\color{black} approach}.

These three examples illustrate the difficulties the instructor may face when she/he  tries to explain real problems, only from the mechanical point of view,  in the framework of the  {\color{black} particle model} \cite{arons89}.
The way to prevent such problem is to avoid the oversimplification introduced by  {\color{black} the particle model when discussing the energetics of the situation}. Of course we may continue to invoke the centre-of-mass to describe the motion of the body, and the concept is even very useful.  Equation (\ref{eq-3}) does apply but one should note that its right hand side must not be confused with work \cite{serwayjewett02},
as mentioned above.

 In thermodynamics, in order to have an energy transfer to or from a system, there should exist so-called reservoirs. This is not surprising when we consider heat flows. In this case we talk about heat reservoirs. For the work one also needs work reservoirs. The atmosphere is a very common work reservoir for $PVT$ systems. It may provide or receive work  without changing its pressure, as the heat reservoir may exchange heat without changing its temperature. By definition, the reservoirs only perform reversible processes. Unlike the heat reservoir, the processes in the work reservoir do not increase the entropy of the universe.

The infinitesimal work for a $PVT$ system is $\delta W = - P \d V$, with $-P$ and $V$ being conjugate variables. Whenever the system interchanges work described by different variables, for instance $mg \, \d h$ as in Joule's paddle experiment \cite{guemez14a}, that is dissipative work, which is equivalent to  heat.

\section{The two blocks example}
\label{sec:twobloksex}

\begin{figure}[htb]
\begin{center}
\hspace*{-0.5cm}
\includegraphics[width=14cm]{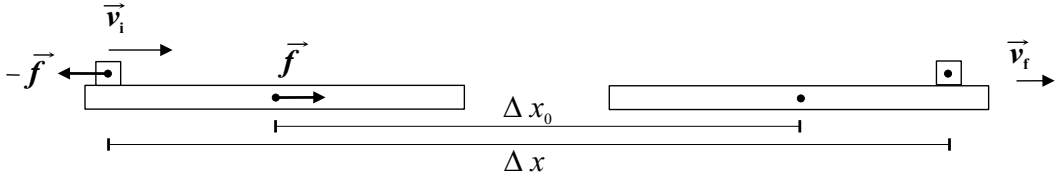}
\end{center}
\vspace*{-0.5cm}
\caption[]{\label{fig1} \small The block at the top, initially with velocity $\vec v_{\rm i}$ moves with friction on top of lower block. After some  time the two blocks move with a common velocity $\vec v_{\rm f}$ The lower block{\color{black}, initially at rest,} moves without friction on top of a horizontal surface not represented in the figure.
}
\end{figure}

One block of mass $m$ and velocity $v_{\rm i}$ gets suddenly on top of another block of higher mass $m_0$ moving horizontally with kinetic friction on its upper surface.
This second block{\color{black}, initially at rest,} may move without friction on top of an infinite horizontal surface. We are interested in a mechanical description of the system as well as in the thermodynamics aspects related to the effect of the friction force. For the sake of simplicity, we consider the friction force as a constant one.  Figure \ref{fig1} represents the situation. After a while the two blocks move together with the common velocity $v_{\rm f}$. {\color{black} Air resistance is neglected.}

The centre-of-mass velocity is equal to the final velocity of the two blocks and it is given by
\beq
V_{\rm cm}= v_{\rm f} = {m \over m + m_0} v_{\rm i}\, .
\eeq
The small block exerts a  force on the lower one, pointing in the positive direction and the lower block exerts an equal and opposite force in the upper block. These contact forces are represented at the centre-of-mass of each block  in figure \ref{fig1}. According to equation (\ref{new0}) applied to each block
\beq
\left\{
\begin{array}{l}
- f  \Delta t = m v_{\rm f} - m v_{\rm i} \\
 \phantom{-} f  \Delta t  = m_0 v_{\rm f}\, ,
 \
 \end{array}
\right.
\label{cvv}
\eeq
and, from equation (\ref{onepai}),
\beq
\left\{
\begin{array}{l}
- f  \Delta x = {1\over 2}m v_{\rm f} ^2 - {1\over 2}m v_{\rm i}^2 \\ \\
 \phantom{.} f  \Delta x_0  = {1\over 2}m_0 v_{\rm f}^2\, .
 \end{array}
\right.
\label{cvf}
\eeq

The variation of the kinetic energy of the system is
\beq
\Delta K = {1\over 2}(m + m_0) v_{\rm f}^2 - {1\over 2}m v_{\rm i}^2 = - {1\over 2}m v_i^2 \left(1\over 1 + m/m_0\right)
\label{cafg1}
\eeq
and this quantity is still equal to the sum of the two equations in (\ref{cvf}), namely
\beq
\Delta K=  f  (\Delta x_0- \Delta x)\, .
\label{cafg2}
\eeq

 {\color{black} We now analyse the situation from the thermodynamical point of view. In the fast, nearly adiabatic process of the collision, there is an internal energy rise of thermal origin (the temperature of the blocks increased). However, the heat transfer to the surrounding during the adiabatic process is negligible.  This fast process is followed by a slow process of thermal conduction from the hot blocks to the cooler air. After this thermalization process the blocks are moving with the velocity $v_{\rm f}$ they had at the end of the fast process, and the increase of internal energy they experienced has been dissipated to the surrounding heat reservoir, with the heat transfer $Q$ equal to the internal energy rise given numerically by equation (\ref{cafg2}), as we shall see.
We can now  apply the first law of thermodynamics, expressed by equation  (\ref{totale}), to the whole process: from the initial state to the final state where the two blocks
are moving with the common velocity $v_{\rm f}$ at at the same temperature, which is equal to their initial temperature and  to the temperature of the heat reservoir, the surrounding air.}
{\color{black} We remind that, in equation  (\ref{totale}), the internal energy, $U$, is the energy of the system is the centre-of-mass frame.}
The variation of the kinetic energy of the centre-of-mass of the system is zero, $\Delta K_{\rm cm}=0$; the variation of the
internal kinetic energy is, therefore, $\Delta K^{\rm int}= \Delta K$; on the other hand, {\color{black}
after thermal equilibrium is reached, the internal energy variation is solely the kinetic energy relative to the center of mass, which has decreased:}
$\Delta U = \Delta K^ {\rm int}$, since the two blocks are at the same temperature in the initial and in the final state {\color{black} (i.e. there is no internal energy variation due to a temperature variation)}; finally, since  $W^{\rm ext}=0$, starting from equation (\ref{totale}) one arrives at
\beq
Q = - {1\over 2}m v_i^2 \left(1\over 1 + m/m_0\right)=  f  (\Delta x_0- \Delta x)\, .
\label{dfgru}
\eeq
This heat transfer to the surrounding is given by a force displacement product which should not be regarded as a work.  The increase of the entropy of the universe   during this process is given by
\beq
\Delta S_{\rm U}= -{Q\over T}= {f  (\Delta x- \Delta x_0) \over T}.
\eeq
{\color{black} where $T$ is the temperature of the reservoir.}

It is interesting to consider the limiting situation of an infinite massive lower block \cite{galili97}.
This limit corresponds to the actual situation of the small  block sliding with friction on top of the ground.  Eventually the body stops, i.e.
$v_{\rm f} =0$. In this case the  force acting on the block is a kinetic friction force \cite{ringlein04},
 usually described by $f=\mu_{\rm k} N$, where $\mu_{\rm k}$ is the kinetic friction coefficient, and $N$ is the normal force exerted by the ground on the
block \cite{besson07}.
In this limiting case, the impulse of the friction force still exists, $m_0 v_{\rm f}\rightarrow \mu_{\rm k}N \, \Delta t$. On the other hand,  in the same limit, the displacement of the lower block vanishes, $\Delta x_0\rightarrow 0$, as well as its kinetic energy,  ${1\over 2} m_0 v_{\rm f}^2 \rightarrow 0$. The displacement of the upper block coincides with the displacement of the centre-of-mass of the system,
$\Delta x \rightarrow \Delta X_{\rm cm}$.
The impulse and the centre-of-mass equations above --- equations (\ref{cvv}) and (\ref{cvf}) --- now reduce to
\beq
\mu_{\rm k} N \, \Delta t = m \, v_{\rm i} \ \ \ \  {\rm and} \ \ \ \ \mu_{\rm k} N \, \Delta X_{\rm cm} = {1 \over 2} m v_{\rm i}^2 \, .
\label{mechasd}
\eeq
From equation (\ref{dfgru}) the heat released in this limiting case can be related both to the initial kinetic energy of the block and to the action of the friction, namely
\beqa
&& - {1\over 2}m v_{\rm i}^2= Q \label{dfpo}\\
&& Q =  -\mu_{\rm k} N \, \Delta X_{\rm cm} \, .
\nonumber
\eeqa

{\color{black} Macroscopic objects, being made of atoms, have many degrees of freedom and can deform, which means that the work done on them may involve forces acting through distances that are different from the displacement of the center of mass, and which can lead to increased internal energy and higher temperature. In fast mechanical processes the heat transfer  (microscopic work of a random nature, in fact) is typically negligible until after the mechanical process has occurred.

The microscopic view of the energy transfers is helpful in the present context. Thus, the energy transfer as heat can described in many situation by an ensemble of thermal photons
and the variation of internal energy related to temperature variations can be described e.g. by phonons in a solid \cite{guemez2016}.}

\subsection{The principle of relativity}
\label{ssec:princrela}

The Galileo's principle of relativity establishes that Newton's equation is covariant under  inertial reference frame transformations. In particular the mass and the force are
frame invariant quantities \cite{tefft07}.
As far as thermodynamics is concerned, the equation expressing the first law (\ref{totale}) is covariant as well. The temperature and internal energy variations and the heat flows are invariant quantities. But, e.g. the kinetic energy and the work are not Galilean invariant. In particular, for the work, because the displacement is not an invariant quantity, but the force is, one expects a change upon a reference from transformation.

 Let us look at the sliding block problem from the point of view of the reference frame S$'$ moving with velocity $V$ with respect to S (the lab frame considered so far) --- for simplicity we analyze the single block problem.

In S$'$, the initial velocity of the block is $v_{\rm i}-V$ and the final velocity is $-V$, hence equations (\ref{mechasd}) read now
\beqa
-\mu_{\rm k} N \, \Delta t &=& -mV-m(v_{\rm i}-V) \\
-\mu_{\rm k} N \, (\Delta X_{\rm cm}-V\, \Delta t) &=& {1 \over 2} m V^2- {1 \over 2} m(v_{\rm i}-V)^2 \, .\nonumber
\eeqa
The first equation is exactly  the first equation (\ref{mechasd}).
The second equation above is
\beq
-\mu_{\rm k} N \, \Delta X_{\rm cm}+\mu_{\rm k} N \, V\Delta t= - {1\over 2} m v_{\rm i}^2 + m v_{\rm i} V \, .
\eeq
Using here the first equation in (\ref{mechasd}) one arrives to the second equation in  (\ref{mechasd}).
This is an expected result because the dynamical information obtained in one inertial reference frame must be equivalent to the information obtained in any other reference frame for the same motion.

Let us finally look at the first law of thermodynamics (\ref{totale}), which now reads $\Delta K'_{\rm cm}+\Delta U'=Q'+W'^{\rm ext}$. {\color{black} We remind that the temperature is the same before and long after the collision, therefore  the internal energy variation is solely due to the internal kinetic energy variation, $\Delta U' =\Delta K'{}^{\rm int}$. On the other hand,} there is an obvious external work related to the friction force due to the velocity of S$'$ with respect to S, namely $-\mu_{\rm k} N \, (-V\, \Delta t)$. The first law of thermodynamics then leads to
\beq
\label{eq:thflspr}
{1 \over 2} m V^2- {1 \over 2} m(v_{\rm i}-V)^2=Q'+\mu_{\rm k} N \, V\, \Delta t\, .
\eeq
After using the first equation (\ref{mechasd}) one concludes that
\beq
Q'=- {1\over 2}m v_{\rm i}^2
\eeq
i.e. $Q'=Q$ [see equations (\ref{dfpo})] and the heat flows are the same in both inertial frames. The external work $\mu_{\rm k} N \, V\, \Delta t$ in (\ref{eq:thflspr})
is performed by the external agent that keeps constant the velocity $-V$ of the floor underneath the block, similarly to the block placed on top of a conveyor belt as discussed in \cite{chabay11}.

\section{The motion of an accelerated car}
\label{sec:motioncar}

\begin{figure}[htb]
\begin{center}
\hspace*{-0.5cm}
\includegraphics[width=10cm]{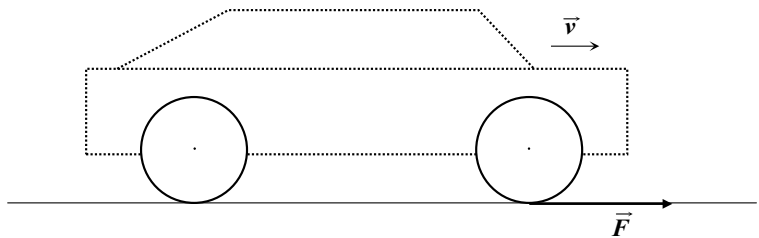}
\end{center}
\vspace*{-0.5cm}
\caption[]{\label{fig2} \small The car, due to the action of force $\vec F$ moves from rest to a final velocity.
}
\end{figure}

The motion of an accelerating car was discussed with some detail in \cite{guemez13b}.
We revisit the example here with some simplifications that allows us to
better concentrate on the main point we want to focus. First of all, we consider a 2-dimensional car. The force on the tractive front wheel, powered
by the engine but exerted by the ground, is represented in figure \ref{fig2} for the car in its accelerating motion.
As shown in \cite{guemez13b} the force exerted on the rear non-tractive wheel
points in the backward direction, but it is rather small and we neglect it here. Moreover, in the energetic discussion in \cite{guemez13b}  we considered the heat exchange with the air and the external work
associated to the functioning engine, to conclude that the appropriate thermodynamical potential is the Gibbs function. However, here we neglect these energy transfers and take
the internal energy variation of the fuel as the relevant thermodynamical potential. This is actually similar to consider the car moving due to the action of an internal source of energy provided by a spiral spring, as it happens in many toys. An additional simplification consists in neglecting the mass of the wheels, therefore their moment of inertia vanishes and there is no kinetic energy associated with the rotation of the wheels.

In the framework of the above mentioned approximations we can write the following equations for the car, of mass $M$, that accelerates in a straight horizontal road.
Let us first consider the motion of the car without air resistance. The force $\vec F$ acts during the time interval $\Delta t$, and the car accelerates from the initial velocity $v_{\rm i}=0$ to the final one $v_{\rm f}=V_{\rm cm}$, traveling a distance $\Delta X_{\rm cm}$. The Newton's second law expressed by (\ref{hjaoi}) and by the centre-of-mass equation (\ref{eq-3}) lead, respectively to
\beq
F \Delta t = M v_{\rm f} - M v_{\rm i}= M \, V_{\rm cm}
\label{ncar1}
\eeq
and
\beq
F \Delta X_{\rm cm}= {1\over 2}M V_{\rm cm}^2=\Delta K_{\rm cm} \, . \label{bhsk}
\label{ncar2}
\eeq
This is also the variation of the kinetic energy of the car, since there is no internal kinetic energies (the wheels are massless):
\beq
\Delta K =  W^{\rm ext} + W^{\rm int} = \Delta K_{\rm cm} \, .\label{senora2}
\eeq
The internal energy variation is solely due to the fuel consumption (by $\xi$ we denote the chemical reaction that transforms fuel in reaction products),
\beq
\Delta U = \Delta U_\xi = n \Delta u_\xi,
\eeq
where $n$ is the number of moles and $\Delta u_\xi<0$ is the molar fuel internal energy variation associated with the chemical reaction.

The tractive force is a static friction force that does not perform any work, $ W^{\rm ext}=0$. Moreover, there is no heat either, $Q=0$, and, as a consequence with no entropy increment $\Delta S_{\rm U}=0$, leading to the conclusion that the tractive force  is non-dissipative. Then, the equation (\ref{totale}) leads to
\beq
{1\over 2}M V_{\rm cm}^2= -n \Delta u_\xi \, .
\eeq
From here and from equation (\ref{senora2}) one concludes that the internal energy variation goes into internal work,
\beq
W^{\rm int} = \Delta K_{\rm cm}
\eeq
which, in turn, by using (\ref{bhsk}), can be related to a force---centre-of-mass displacement product, namely
\beq
-n \Delta u_\xi=F\,  \Delta X_{\rm cm}\, ,
\eeq
where the indicated displacement is not the displacement of the application point of the force (such displacement is zero).

Before discussing the effect of the air drag force, it is instructive to simulate its effect using a conservative force opposite to $\vec F$.
This conservative force can be an electric charge, $q$, placed at the centre-of-mass of the car and assuming that the car is moving in a region of a constant electric field, $\vec E$. The electric force is given by $F_E=-qE$, the negative sign indicating that the force is opposite to the direction of the motion. Due to this new force, the car, after the same time interval, $\Delta t$, only reaches the velocity $\overline V_{\rm cm}<V_{\rm cm}$ after traveling a distance $\overline X_{\rm cm}<X_{\rm cm}$. The dynamical equations (\ref{ncar1}) and (\ref{ncar2}) now read
\beq
(F-qE) \, \Delta t = M \overline V_{\rm cm} \ \ \ \ {\rm and} \ \ \ \ \ (F-qE)\overline X_{\rm cm} = {1 \over 2}  M \overline V_{\rm cm}^2 = \Delta \overline K_{\rm cm}\, .
\eeq
There is an external work, $W^{\rm ext} = -q\, E\, \overline X_{\rm cm}$, so the variation of the system's kinetic energy, equation (\ref{senhora}), is
\beq
\Delta \overline K =-q\, E\, \overline X_{\rm cm}+ W^{\rm int}={1 \over 2}   M \overline V_{\rm cm}^2=\Delta \overline K_{\rm cm}\, .
\label{eqklinha}
\eeq
Note that $ W^{\rm int}$ is the same as before because the ca´r's engine operates exactly in the same regime during exactly the same time interval. Therefore, $\Delta U = n \Delta u_\xi$ and $ W^{\rm int}=-n \Delta u_\xi$ as before the introduction of the electric  braking force. The first law of thermodynamics leads to
\beq
{1 \over 2}  M \overline V_{\rm cm}^2+n \Delta u_\xi=-q\, E\, \overline X_{\rm cm} + Q\, .
\eeq
Using (\ref{eqklinha}) one concludes that $Q=0$ and, as a consequence, $\Delta S_{\rm U}=0$.
The energy transfer to the car due to the electric force is ultimately due to an external work reservoir (for instance, a large capacitor creating the electric field). This external work can still be written as
\beq
W^{\rm ext} = -q\, E\, \overline X_{\rm cm}= {1 \over 2} M   \overline V_{\rm cm}^2 - {1 \over 2} M   V_{\rm cm}^2
\eeq
and this energy can be totaly recovered since the difference of these kinetic energies (with and without electric force) is stored elsewhere as electric potential energy.

Let us now assume that the braking electric force is replaced by a  {\color{black} drag force}  of equal magnitude $F_E \rightarrow f$.
In this case, the dynamical discussion remains the same just with this replacement: $qE=f$. The final velocity is the same, $\overline V_{\rm cm}$;  the distance traveled is the same, $\overline X_{\rm cm}$; the variation of internal energy, the work of the internal forces is the same, $\Delta U=n\Delta u_\xi=-W^{\rm int}$.
For completeness, we rewrite the above equations for this new situation as
\beq
(F-f) \, \Delta t = M \overline V_{\rm cm} \ \ \ \ {\rm and}\ \ \ \ \ (F-f)\overline X_{\rm cm} = {1 \over 2}  M \overline V_{\rm cm}^2 = \Delta \overline K_{\rm cm}
\eeq
and
\beq
\Delta \overline K=-f \, \overline X_{\rm cm}+ W^{\rm int}={1 \over 2}   M \overline V_{\rm cm}^2=\Delta \overline K_{\rm cm} \, ,
\label{senhora3}
\eeq
where the ``mechanical" effect of the drag has been included in (\ref{senhora3}).
 The same term, $-f \, \overline X_{\rm cm}$, should enter on the right hand side of the expression of the first law (\ref{totale}), accounting for the energy exchange between the car and the surrounding air:
\beq
{1 \over 2}   M \overline V_{\rm cm}^2-n\Delta u_\xi= -f \, \overline X_{\rm cm}
\eeq
Ultimately, the air resistance force is the time/space-averaged momentum transfer rate of random collisions with air molecules, in which purely conservative electric forces act microscopically. But macroscopically this force performs dissipative work equivalent to heat
\beq
Q=-f\, \Delta X_{\rm cm}<0\, ,
\eeq
which eventually flows to the surrounding air, the heat reservoir, {\color{black}
leading to the entropy increase $\Delta S_{\rm U} >0$. The kinetic energy obtained by the car results from an internal work reservoir. But,
from the thermodynamical point of view, the effect of the drag force cannot be associated to a work
reservoir.}

The previous expression can still be written as
\beq
Q={1 \over 2}   M \overline V_{\rm cm}^2-{1 \over 2}   M V_{\rm cm}^2\, .
\eeq

The difference in the kinetic energies with and without air resistance is equal to the energy released at heat, leading to an increase of the entropy of the universe. Since there is no potential behind the drag force, the energy involved due to the action of that force cannot get organized as a potential energy. Its usefulness has been lost.

\section{Sliding and rotating disc in a slippery slope}
\begin{figure}[htb]
\begin{center}
\hspace*{-0.5cm}
\includegraphics[width=7cm]{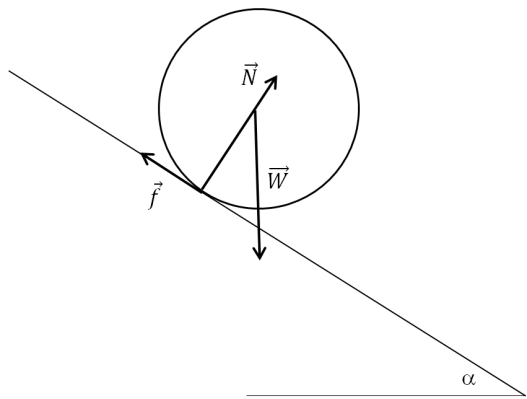}
\end{center}
\vspace*{-0.5cm}
\caption[]{\label{fig3} \small Forces applied to a rotating disc descending an incline (with or without slipping).
}
\end{figure}

In this section we study a sliding and rotating disc, of mass $M$ and moment of inertia $I={1 \over 2} M R^2$, where $R$ is its radius, descending a slippery incline, whose inclination angle is $\alpha$. The kinetic friction force is $f=\mu_{\rm k}N= \mu_{\rm k}Mg \cos \alpha$ (see figure \ref{fig3}) and the non-slipping condition is not observed: $\omega R \not= V_{\rm cm}$, where $V_{\rm cm}$ is the instantaneous centre-of-mass velocity and $\omega$ is the instantaneous angular velocity around the rotation axis. \

For the translational motion, Newton's equation (\ref{hjaoi}) leads to
\beq
(M\, g\sin \alpha - \mu_{\rm k} Mg \cos \alpha) \, \Delta t = M  V_{\rm cm}\, .
\eeq
where $V_{\rm cm}$ represents now the final centre-of-mass velocity i.e. at the base of the incline (zero initial velocity is assumed) and the distance $\Delta X_{\rm cm}$  was traveled in the time interval $\Delta t$.
The centre-of-mass equation (\ref{eq-3}) leads to
\beq
(Mg\sin \alpha -\mu_{\rm k} Mg \cos \alpha) \Delta X_{\rm cm} = {1 \over 2} M V_{\rm cm}^2\, .
\eeq
After integration, the rotational equations (\ref{rot1}) and (\ref{rot2}) lead to
\beq
I \omega = \mu_{\rm k} MgR\Delta t\, ,
\eeq
where $\omega$ now stands for the final angular velocity, and
\beq
{1 \over 2} I \omega^2 = \mu_{\rm k} MgR\theta_0
\eeq
(by $\theta_0$ we denote the total rotation angle of the disc around its axis).

{\color{black} Regarding the thermodynamical discussion, to be realistic, one has to admit that there is a temperature increase of the disc due to friction, and this gives rise to an energy transfer as heat to the
surrounding air, the heat reservoir. The corresponding increment of the internal energy is $\Delta U_{T}$. Once a steady state is reached the disc moves at a constant temperature that is sufficiently higher than the air temperature for that value of the rate of $Q$ to go to the air that keeps the disc temperature constant.}

The internal energy variation, {\color{black} $\Delta U= {1 \over 2} I \omega^2+ \Delta U_T$,} is  due to the rotational energy variation and {\color{black} to the temperature increment of the disc}. The  external work of the conservative forces is $W^{\rm ext}=Mg \Delta X_{\rm cm} \sin \alpha$.   {\color{black} The work of the dissipative force leads to the increment $\Delta U_T$ of the internal energy and ultimately to an heat  transfer to the surrounding air:
\beq
{1 \over 2}M V^2_{\rm cm} + {1\over 2} I \omega^2 + \Delta U_T= Mg \Delta X_{\rm cm} \sin \alpha + Q\, .
\eeq
}
The consistency of the above equations requires {\color{black}
\beq
Q = -\mu_{\rm k} Mg \cos \alpha (\Delta X_{\rm cm}- R\theta_0)+ \Delta U_T <0
\eeq
i.e. there should be an heat transfer to the surrounding. The lower the internal energy increase of the disc, the higher this heat transfer will be.}
 {\color{black} The increment of the entropy of the universe should be positive $\Delta S_{\rm U}>0 $.}
 Should the rolling without slipping condition be observed, $\Delta X_{\rm cm}= R\theta_0$, that term would be zero, {\color{black} there will be no increment of the internal energy due to thermal effects ($\Delta U_T=0$)} and the heat would be zero as well.

\color{black}
\section{Conclusions}
\label{sec:conclus}

We discussed the effects of dissipative forces acting on a system.
One effect of the dissipative force is the reduction of the energy of the system due to a heat transfer to the surrounding. This heat may be  quantitatively given as the product of a force by a characteristic displacement of the body, being accounted for as pseudo-work in Newton's second law.

As a result of the examples discussed in this article, we may argue, in general, that an energy exchange between the system and the surrounding expressed in terms of variables that are not state variables of the system, is dissipative work, thermodynamically equivalent to heat. In other words,  when the energy transfer cannot be expressed by products of conjugate state variables of the  system, that energy is not macroscopically organized and therefore it cannot be stored and recovered as work.

{\color{black} At a mesoscopic scale, friction and drag forces are ultimately conservative electric forces that do work to deform multiparticle systems, with an associated temperature rise, which leads to a heat transfer to the surroundings. At a macroscopic scale, from the mechanical point of view, these forces perform dissipative work.}

We stress again that instructors should be aware of the limitations introduced when an extensive body acted upon by dissipative forces is reduced to a point like object: in that case one incorrectly refers to the dissipative force---centre-of-mass displacement product as work, though actually it is pseudo-work \cite{arons99}.

\vspace*{1cm}
{\color{black}
We are very grateful to an anonymous reviewer for the very useful suggestions that helped to considerably improve the article.
}

\end{document}